\documentclass[fleqn,10pt]{wlscirep}
\usepackage[utf8]{inputenc}
\usepackage[T1]{fontenc}
\title{The theoretical direct-band-gap optical gain of  Germanium nanowires}

\author[1,2]{Wen Xiong}
\author[3]{Jian-Wei Wang}
\author[1,*]{Wei-Jun Fan}
\author[1]{Zhi-Gang Song}
\author[1]{Chuan-Seng Tan}
\affil[1]{Nanyang Technological University, School of EEE,  50 Nanyang Avenue, Singapore, 639798, Singapore}
\affil[2]{Chongqing University, School of Physics, No. 55 South Road, University Town, Chongqing, 401331, People's Republic of China}
\affil[3]{University of Electronic Science and Technology of China, School of Materials and Energy,  No. 4, Section 2, North Jianshe Road, Chengdu, 610054, People's Republic of China}

\affil[*]{ewjfan@ntu.edu.sg}

\begin{abstract}
We calculate the electronic structures of Germanium nanowires by taking the effective-mass theory. The electron and hole states at the $\Gamma$-valley are studied via the eight-band $\textbf{k.p}$ theory. For the [111] $L$-valley, we expand the envelope wave function using Bessel functions to calculate the energies of the electron states for the first time. The results show that the energy dispersion curves of electron states at the $L$-valley are almost parabolic irrespective of the radius of Germanium nanowires. Based on the electronic structures, the density of states of Germanium nanowires are also obtained, and we find that the conduction band density of states mostly come from the electron states at the $L$-valley because of the eight equivalent degenerate $L$ points in Germanium. Furthermore, the optical gain spectra of Germanium nanowires are investigated. The calculations show that there are no optical gain along $z$ direction even though the injected carrier density is 4$\times$$10^{19}$ cm$^{-3}$ when the doping concentration is zero, and a remarkable optical gain can be obtained when the injected carrier density is close to 1$\times$$10^{20}$ cm$^{-3}$, since a large amount of electrons will prefer to occupy the low-energy $L$-valley. In this case, the negative optical gain will be encountered considering free-carrier absorption loss as the increase of the diameter. We also investigate the optical gain along $z$ direction as functions of the doping concentration and injected carrier density for the doped Germanium nanowires. When taking into account free-carrier absorption loss, the calculated results show that a positive net peak gain is most likely to occur in the heavily doped nanowires with smaller diameters. Our theoretical studies are valuable in providing a guidance for the applications of Germanium nanowires in the field of microelectronics and optoelectronics. 
\end{abstract}
\begin{document}

\flushbottom
\maketitle
\thispagestyle{empty}

\section*{Introduction}

Group IV semiconductor nanowires have stimulated a lot of research interests owing to their broad range of applications in the fields of nanoelectronics and optoelectronics. As we know, Gemanium (Ge) is a special kind of semiconductor material compared with Silicon (Si). Because on the one hand, it has small carrier effective mass, indicating that the large carrier mobilities in the devices made of Ge material, and on the other hand, a small direct band gap is about 0.80 eV at the minimum $\Gamma$-vally, which is 136 meV larger than the indirect band gap at the minimum $L$-valley\cite{1}. The lower energy in the $L$-valley makes the applications of Ge-based optical devices in laser fields a big obstacle, because most electrons will fill the indirect $L$-valley firstly and only a small amount of electrons can leak into the direct $\Gamma$-valley, thus the low luminescent efficiency will be encountered. 

Difficulties caused by the indirect-band-gap nature of Ge material have forced researchers to find solutions. In the past decades, various efforts including heavy n-type doping\cite{2,3,4}, tensile strain\cite{5,6,7,8} and incorporating tin (Sn)\cite{9,10,11,12,13} have been implemented to adjust the band structures of bulk or low-dimensional Ge material. No matter what kind of methods, the goal is to diminish the energy difference between the $L$-valley and $\Gamma$-valley, and to enhance the occupied proportion of the electrons at the $\Gamma$-valley. So far, for the nanostructures of Ge, a lot of research work mainly foucs on the mechanical, electrical and optical properties of Ge or Ge alloy quantum wells (QWs), such as Ge\cite{14}, Ge/GeSi\cite{15}, Si/GeSiSn\cite{16}, and GeSn/GeSiSn\cite{17} QWs, and some of these work have been proved effectively to reverse the minimum of $\Gamma$-valley and $L$-valley. Besides the quantum wells, a number of other papers are devoted to studying the properties of Ge-related nanowires. For example, in experiments, Ge nanowires were reported to be synthesized successfully in many groups\cite{1,18,19,20}, and especially, the excellent size control of the diameters of Ge nanowires below 10 nm were able to be realized\cite{19,21,22}. In comparison with the bulk structures, there are many advantages for the use of nanowires or quantum wires in the field of optoelectronics. Firstly, the carriers are confined in two dimensions and only free in one direction, which may lead to a higher spontaneous emission rate. Secondly, nanowires can withstand several times of mechanical strain, and the related experimental results on Ge nanowires have been reported\cite{23}. Finally, nanowires can be easily integrated in photonic-integrated circuits because of the small size, as reported in Ref. \cite{1}. Parallel to the experimental work, the theoretical work on Ge nanowires were also carried out. The electronic properties of Ge or Ge/Si nanowires with and without the strain effects were mainly calculated via the first-principle method\cite{24,25}. However, there are a lot of drawbacks using this method to study nanowires. On the one hand, it is time-consuming or impossible for calculating the electronic structures of nanowires with large diameters, thus the diameters of the calculated nanowires in Refs. \cite{24,25} did not exceed 6 nm. On the other hand, the detailed physics cannot be clarified clearly when calculating the optical properties of nanowires. For example, similar to the quantum wells in the field of lasers, nanowires can also be used as miniature lasers\cite{26,27,28}, while the optical gain of nanowires can not be calculated and analyzed in detail by \textit{ab initio} method.

In this paper, we establish the direct-band-gap optial gain theory of Germanium nanowires. We take the effective-mass theory to investigate the electronic structures and optical gain of Ge nanowires with the diameters from 6 nm to 18 nm, and the indirect [111] $L$-valley is taken into account. At the $\Gamma$-valley, the eight-band $\textbf{k.p}$ theory is used to calculate the enegies of the electron and hole states. Meanwhile, the main focus is on the [111] $L$-valley of Ge nanowires. For the first time, we adopt the Bessel functions to expand the envelope wave functions of the electrons at the $L$-valley, then the electron states at the $L$-valley can be obtained. Further, the optical gain spectra as functions of the diameters, doping concentrations and injected carrier densities are calculated, and free-carrier absorption (FCA) loss is also considered. The followings are the organization of this paper. First of all, the theoretical model are presented, including the computational models of the electronic structures, density of states, optical gain and FCA loss of Ge nanowires. Secondly, we give the corresponding numerical results and discussions. Finally, a brief summary about our theoretical results are drawn.

\section*{The theoretical model}

\subsection*{The electronic structures of Ge nanowires at the $\Gamma$-valley and $L$-valley}

At the $\Gamma$-valley, based on the eight Bloch basis: $|S\rangle\uparrow$, $\frac{1}{\sqrt{2}}(|X\rangle+i|Y\rangle)\uparrow$, $|Z\rangle\uparrow$,
$\frac{1}{\sqrt{2}}(|X\rangle-i|Y\rangle)\uparrow$ and $|S\rangle\downarrow$, $\frac{1}{\sqrt{2}}(|X\rangle+i|Y\rangle)\downarrow$, $|Z\rangle\downarrow$,
$\frac{1}{\sqrt{2}}(|X\rangle-i|Y\rangle)\downarrow$, the Hamiltonian of zinc-blende structure Ge without the stress can be expressed as follows\cite{29}:
\begin{equation}
H_{eight}=\left(
            \begin{array}{cc}
              H_{int}^{U} & 0 \\
              0 & H_{int}^{L} \\
            \end{array}
          \right)+H_{so}
\end{equation}
where $H_{int}^{U}$=$H_{int}^{L}$ is a 4$\times$4 matrix, which can be written as:
\begin{equation}
H_{int}^{U}=\frac{1}{2m_{0}}\left(
              \begin{array}{cccc}
                \epsilon_{g}+P_{e} & \frac{i}{\sqrt{2}}p_{0}p_{+} & ip_{0}p_{z} & \frac{i}{\sqrt{2}}p_{0}p_{-} \\
                -\frac{i}{\sqrt{2}}p_{0}p_{-} & -P_{1} & -S & -T \\
                -ip_{o}p_{z} & -S^{*} & -P_{3} & -S \\
                -\frac{i}{\sqrt{2}}p_{0}p_{+} & -T^{*} & -S^{*} & -P_{1} \\
              \end{array}
            \right)
\end{equation}
In the above matrix, the elements $P_e$, $P_1$, $P_3$, $T$, $T^*$, $S$ and $S^*$ can be found in Ref. \cite{29} and it should be noticed that the parameters $\alpha$, $m_c$, $L_p$, $M$, $N$, $E_p$, $E_g$ and $\Delta_{so}$ are used to determine the band structures of Ge nanowires at the $\Gamma$-valley. $H_{so}$ is the valence band spin-orbit coupling (SOC) Hamiltonian, which is an 8$\times$8 sparse matrix. There are only following non-zero matrix elements in $H_{so}$, namely, $H_{so}(3,3)$=$H_{so}(7,7)$=-$\Delta_{so}$/3, $H_{so}(4,4)$=$H_{so}(6,6)$=2$H_{so}(3,3)$, $H_{so}(3,6)$=$H_{so}(6,3)$=$\sqrt{2}$$H_{so}(3,3)$ and $H_{so}(4,7)$=$H_{so}(7,4)$=-$\sqrt{2}$$H_{so}(3,3)$. We can use the Bessel expansion method to solve the above Hamiltonian\cite{29}, and then the electron and hole states at the $\Gamma$-valley can be obtained. 

For indirect-band-gap semiconductor Ge, the electron states of the [111] $L$-valley must be considered, because the band gap at the $L$-valley is only a little smaller than that of the $\Gamma$-valley. As we know, there are eight degenerate equivalent $L$ points in Ge, whereas only half of the ellipsoids are inside the first Brillouin zone. Hence, for simplicity, we only need to take into account one particular $L$-valley for [001] Ge nanowires, which is the same as the case of Ge-related quantum wells (QWs)\cite{17}. The Hamiltonian of the [111] $L$-valley of [001] Ge nanowires can be expressed as\cite{17}
\begin{equation}
H_L^{[111]}=\frac{\hbar^2}{2m_1^*}(k_z-\pi/a)^2+\frac{\sqrt{2}\hbar^2}{3m_2^*}k_1(k_z-\pi/a)+\frac{\hbar^2}{2m_3^*}k_1^2+\frac{\hbar^2}{2m_4^*}k_2^2+V_{L}(r,\theta)
\end{equation}
where $1/m_1^*$=$1/(3m_{l,L})$+$2/(3m_{t,L})$, $1/{m_2^*}$=$1/m_{l,L}$-$1/m_{t,L}$, $1/m_3^*$=$2/(3m_{l,L})$+$1/(3m_{t,L})$, $m_4^*$=$m_{t,L}$, and $m_{l,L}^*$ and $m_{t,L}^*$ are the longitudinal and transverse effective masses at the $L$-valley. $k_1$=$(k_x+k_y)$$/\sqrt{2}$ and $k_2$=$(-k_x+k_y)$$/\sqrt{2}$. $k_z$ is a good quantum number for Ge nanowires, which is relative to $\pi/a$, namely, the minimum point of the $L$-valley, where $a$ is the lattice constant of Ge. $V_L(r,\theta)$ is the confined potential of the electron at the $L$-valley in Ge nanowires, which can be taken as zero in the nanowires and infinity outside the nanowires, respectively. Then, the wave function of the electron at the $L$-valley can be written as 
$\chi^{[111]}_L$=$F_{L}(r,\theta,k_z)$$\upsilon_L$, where $F_{L}(r,\theta,k_z)$ and $\upsilon_L$ are the envelope wave function and Bloch states at the $L$-valley, respectively. In order to obtain the energies of the electron states at the $L$-valley, we must solve the following Schr\"{o}dinger equation
\begin{equation}
H_L^{[111]}F_{L}(r,\theta,k_z)=E_{n_c}^{L}(k)F_{L}(r,\theta,k_z)
\end{equation}
Because the cylindrical symmetry of [001] Ge nanowires has been broken at the [111] $L$-valley, which is different from the case of the $\Gamma$-valley. We must expand the envelope wave function $F_{L}(r,\theta,k_z)$ as
\begin{equation}
F_{L}(r,\theta,k_z)=\sum_{n,m}a_{n,m}A_{n,m}j_m(k_n^{m}r)e^{im\theta}e^{ik_zz}
\end{equation}
where $j_{m}(k_{n}^{m}r)$ is the $m$th order Bessel function, which ensures that the wave function of the electron at the boundary $r$=$R$ is zero at the [111] $L$-valley. In addition, $k_{n}^{m}$=$\frac{\alpha_{n}^m}{R}$, and $R$ is the radius of Ge nanowires. The parameter $\alpha_n^m$ is the $n$th zero point of $m$ order Bessel function, and $A_{n,m}$ is the normalization constant of $j_{m}(k_{n}^{m}r)$. The detailed matrix elements of the above Eq. (5) are presented in Appendix and we can obtain the energy level of each electron state at the $L$-valley by diagonalizing the Hamiltonian $H_L^{[111]}$ numerically.  

\subsection*{The density of states and quasi-Fermi levels of conduction and valence band of Ge nanowires}

Based on the electronic structures of Ge nanowires at the $\Gamma$-valley and $L$-valley, the conduction and valence band density of states (DOS) of Ge nanowires can be calculated as
\begin{equation}
D_c(E)=\frac{1}{2\pi S}\sum_{n_c}\int_{BZ}\delta[E-E_{n_c}^{\Gamma}(k_z)]dk_z+8\frac{1}{2\pi S}\sum_{n_c}\int_{BZ}\delta[E-E_{n_c}^{L}(k_z)]dk_z
\end{equation}
\begin{equation}
D_v(E)=\frac{1}{2\pi S}\sum_{n_v}\int_{BZ}\delta[E-E_{n_v}^{\Gamma}(k_z)]dk_z
\end{equation}
where $E_{n_c}^{\Gamma}(k_z)$, $E_{n_c}^{L}(k_z)$ are the energies of the $n_c$th electron state at the $\Gamma$-valley and $L$-valley, respectively, and $E_{n_v}^{\Gamma}(k_z)$ is the energy of the $n_v$th hole state at the $\Gamma$-valley. $S$ is the cross-section area of the nanowires. A factor of 8 in Eq. (6) means that the spin degeneracy of four equivalent $L$-valley is included in $D_c(E)$ of the conduction band, and it should be noticed that the spin of the electron and hole states at the $\Gamma$-valley has also been included in Eqs. (6) and (7).    

We can determine the quasi-Fermi levels $E_{fc}$ of the conduction band numerically at a given temperature $T$ using $N_e$=$N_{d}$+$\Delta n$, where $N_d$ and $\Delta n$ are the donor doping concentration and injected carrier density, respectively. Because Ge nanowires should be heavily n-type doped, the donor doping concentration $N_d$ is much larger than that of the acceptor doping concentration $N_a$, and $N_a$ can be neglected. Therefore, $E_{fc}$ can be calculated via the following equation 
\begin{equation}
\begin{aligned}
N_e= N_d+\Delta n=& \sum_{n_c}\frac{1}{2\pi S}\int_{BZ}\Big[\exp\big(\frac{E_{n_c}^{\Gamma}(k_z)-E_{fc}}{k_{B}T}\big)+1\Big]^{-1}dk_z+ \\ & 8\sum_{n_c}\frac{1}{2\pi S}\int_{BZ}\Big[\exp\big(\frac{E_{n_c}^{L}(k_z)-E_{fc}}{k_{B}T}\big)+1\Big]^{-1}dk_z
\end{aligned}
\end{equation}
where $k_B$ is the Boltzmann's constant. Similarly, the quasi-Fermi levels $E_{fv}$ of the valence band can be calculated as
\begin{equation}
N_h=\Delta n=\sum_{n_v}\frac{1}{2\pi S}\int_{BZ}\Big[\exp\Big(\frac{E_{fv}-E_{n_v}^{\Gamma}(k_z)}{k_{B}T}\Big)+1\Big]^{-1}dk_z
\end{equation}

\subsection*{The optical gain of Ge nanowires}

In the previous subsection, $E_{fc}$ and $E_{fv}$ in Eqs. (8) and (9) are determined numerically, and after that we can calculate the optical gain $g(E)$ of Ge nanowires via the following equation
\begin{equation}
\begin{aligned}
g(E)= & \Big[1-\exp\big(\frac{E-\Delta F}{k_{B}T}\big)\Big]\frac{\pi {e}^2\hbar}{m_0^{2}\varepsilon_{0}n_{r}cE}\times \\ &
\sum_{n_c,n_v}\int_{BZ}\frac{dk_z}{2\pi S}|M_{n_c,n_v}|^2 f_{n_c}(k_z)\big(1-f_{n_v}(k_z)\big)\frac{1}{\pi}\frac{\hbar/\tau}{(E_{eh}-E)^2+(\hbar/\tau)^2}
\end{aligned}
\end{equation}
where $m_0$ and $e$ are the mass the charge of the free electron, respectively, $\varepsilon_{0}$ is the dielectric constant of the vacuum, and $n_r$ is the refractive index of Ge material. $\Delta F$=$E_{f_c}$-$E_{f_v}$ is the separation between the quasi-Fermi levels of $E_{f_c}$ and $E_{f_v}$, $E_{eh}$ is the transition energy, $E$ is the photon energy and $\tau$ is the relaxation time. $f_{n_c}(k_z)$ and 1-$f_{n_v}(k_z)$ are the Fermi-Dirac distribution functions of the electron and hole, respectively, which can be expressed as 
\begin{equation}
f_{n_c}^{\Gamma}(k_z)=\Big[\exp\big(\frac{E_{n_c}^{\Gamma}(k_z)-E_{fc}}{k_{B}T}\big)+1\Big]^{-1}, ~~~ f_{n_c}^{L}(k_z)=\Big[\exp\big(\frac{E_{n_c}^{L}(k_z)-E_{fc}}{k_{B}T}\big)+1\Big]^{-1}
\end{equation}
\begin{equation}
1-f_{n_v}(k_z)=\Big[\exp\big(\frac{E_{fv}-E_{n_v}^{\Gamma}(k_z)}{k_{B}T}\big)+1\Big]^{-1}
\end{equation}
$|M_{n_c,n_v}|^2$ is the squared optical transition matrix element from the $n_c$th electron state to $n_v$th hole state. If the optical transition selection rule of nanowires is satisfied, $|M_{n_c,n_v}|^2$ along $z$ direction can be expressed as\cite{29}
\begin{equation}
\Big|M_{n_c,n_v}^z\Big|^2=\Big|\sum_{n}(e_{m,k_z,n,\uparrow}^{*}c_{m,k_z,n,\uparrow}+e_{m+1,k_z,n,\downarrow}^{*}c_{m+1,k_z,n,\downarrow})\Big|^2\Big|\langle S|\hat{p}_z|Z\rangle\Big|^2
\end{equation}
where $\langle S|\hat{p}_z|Z\rangle$ relates to the Kane energy $E_p$\cite{30}. $|M_{n_c,n_v}|^2$ along $x$ direction can also be calculated similarly. 

\subsection*{Free-carrier absorption loss of Ge nanowires}

As we know, the free-carrier absorption (FCA) loss of heavily doped Ge nanowires can not be neglected. By using Drude-Lorentz equation, FCA loss can be calculated as\cite{16}
\begin{eqnarray}
\alpha_f=\frac{e^{3}\lambda^{2}}{4\pi^{2}c^{3}\epsilon_{0}n_{r}}\Big[\frac{N_e^{\Gamma}}{\mu_{\Gamma}(m_{c\Gamma}^*)^{2}}+\frac{N_e^{L}}{\mu_{L}(m_{cL}^*)^{2}}+\frac{N_h}{\mu_{h}(m_{h}^*)^{2}}\Big]
\end{eqnarray}
where the parameters $e$, $c$, $\epsilon_{0}$ and $n_r$ are the same as those in Eq. (10), and $\lambda$ is the free space wavelength. $m_{c\Gamma}^*$ and $m_{cL}^*$ are the conductivity effective masses at the $\Gamma$-valley and $L$-valley, respectively. $m_{c\Gamma}^*$ can be considered as $m_c$ due to the isotropy at the $\Gamma$-valley, and $m_{cL}^*$ can be calculated as follows using two effective-mass $m_{l,L}^*$ and $m_{t,L}^*$ at the $L$-valley
\begin{eqnarray}
m_{cL}^*=\frac{3}{\frac{1}{m_{l,L}^*}+\frac{2}{m_{t,L}^*}}
\end{eqnarray}
For unstrained Ge, the top valence band is heavy hole, thus we can use the heavy hole effective-mass as $m_h^*$, namely, $m_h^*$=$1/(\gamma_1-2\gamma_2)$\cite{31}. $N_e^{\Gamma}$ and $N_e^{L}$ are the electron concentrations at the $\Gamma$-valley and $L$-valley, respectively. $\mu_{\Gamma}$ and $\mu_{L}$ are the electron mobilities at the $\Gamma$-valley and $L$-valley, respectively. $\mu_{L}$ can be expressed as\cite{16}
\begin{eqnarray}
\mu_{L}=\frac{3900}{1+\sqrt{N_e^L\times10^{-17}}}
\end{eqnarray}
where the units of $\mu_{L}$ and $N_e^L$ are cm$^2$ V$^{-1}$ s$^{-1}$ and cm$^{-3}$, respectively. If we assume the scattering times of the electrons at the $\Gamma$-valley and $L$-valley are the same, $\mu_{\Gamma}$ can be calculated via the relation $\mu_{\Gamma}$=$\frac{m_{cL}^*}{m_{c\Gamma}^*}$$\mu_L$. In additon, by fitting the experimental data, we can calculate the hole mobility $\mu_{h}$ of Ge as\cite{16}
\begin{eqnarray}
\mu_{h}=\frac{1900}{1+\sqrt{N_{h}\times2.1\times10^{-17}}}
\end{eqnarray} 
where the units of $\mu_{h}$ and $N_{h}$ are also cm$^2$ V$^{-1}$ s$^{-1}$ and cm$^{-3}$, respectively.

\section*{Numerical results and discussions}

\subsection*{The electron and hole states of Ge nanowires}

In this section, first of all, we will calculate the electronic structures of Ge nanowires. During the calculation of the band structures at the $\Gamma$-valley, the parameters of Ge material at $T$=300 K are taken: three Luttinger's parameters $\gamma_1$=13.38, $\gamma_2$=4.24, $\gamma_3$=5.69\cite{17,32}, the effective mass of the electron at the $\Gamma$-valley $m_c$=0.038 $m_0$\cite{17,32}, the spin-orbit splitting energy $\Delta_{so}$=0.29 eV\cite{17}, the band gap $E_{g}^{\Gamma}$=0.7985 eV\cite{17}. The Luttinger's parameters can be converted to our used parameters via the following relations\cite{33}
\begin{equation}
L_p=\gamma_1+4\gamma_2,~~~ M=\gamma_1-2\gamma_2,~~~N=6\gamma_3    
\end{equation}    
In order to eliminate the spurious solutions, we can set the Kane energy $E_p$ suggested by Foreman as\cite{33}
\begin{equation}
E_p=\frac{3m_0}{m_c}\frac{1}{2/E_{g}^{\Gamma}+1/(E_{g}^{\Gamma}+\Delta_{so})}
\end{equation}
While the parameters of Ge material at the $L$-valley are taken: two effective masses $m_{l,L}$=1.57 $m_0$ and $m_{t,L}$=0.0807 $m_0$\cite{17}, the lattice constant $a$=5.6573 {\AA}\cite{17}, and the band gap $E_{g}^{L}$=0.664 eV\cite{17}. 

Figures 1(a) and 1(b) show the six lowest electron states and ten highest hole states of Ge nanowires with the diameter $D$=6 nm at the $\Gamma$-valley, respectively. Because of the cylindrical symmetry of [001] Ge nanowires at the $\Gamma$-valley, the total angular momentum projection along $z$ axis of nanowires $J$ is a good quantum number, and each electron and hole state is doubly degenerate with the same $|J|$. Therefore, we can use $e_n^J$ and $h_n^J$ to denote each electron state and hole state at the $\Gamma$-valley conveniently, where the meanings of $n$ and $J$ can be found in Ref. \cite{29}. In figure 1(a), we can see that the ground electron state is $e_0^{1/2}$ at the $\Gamma$ point, and $e_0^{1/2}$ is degenerate with $e_0^{-1/2}$. According to the analysis, it is found that $e_0^{1/2}$ is mainly composed by $|S\rangle$$\uparrow$ Bloch basis component, which can also be seen in the probability density of figure 2(a). The reason is that the zero order Bessel function $j_0(r)$ is the largest at $r$=0, the probability density in figure 2(a) is large near the radial center $r$=0. The two following electron states are $e_1^{1/2}$ and $e_0^{3/2}$ at the $\Gamma$ point, respectively, and these two states are close at the $\Gamma$ point. As the increase of the wave vector $k_z$, $e_0^{1/2}$, $e_1^{1/2}$ and $e_0^{3/2}$ will cross, which means that the electron states at the $\Gamma$-valley present a slight non-parabolic band behaviour. The probability densities of $e_1^{1/2}$ and $e_0^{3/2}$ at the $\Gamma$ point are also presented in figures 2(b) and 2(c), which are very different from that of the ground state $e_0^{1/2}$. In figure 1(b), it can be seen that the three highest hole states are $h_0^{1/2}$, $h_1^{1/2}$ and $h_0^{3/2}$ at the $\Gamma$ point, respectively. The main Bloch basis component is $|Z\rangle$$\uparrow$ in $h_0^{1/2}$, which accounts for about 80\% of the total component. Similar to the probability density of $e_0^{1/2}$ in figure 2(a), we can also analyze the main component from the probability density of $h_0^{1/2}$ in figure 2(e). $h_1^{1/2}$ is mainly composed by $\frac{1}{\sqrt{2}}(|X\rangle+i|Y\rangle)\uparrow$ and  $\frac{1}{\sqrt{2}}(|X\rangle-i|Y\rangle)\uparrow$ Bloch basis components, which accounts for about 80\% of the total component. Unlike the electron states, it can be seen that the hole states show strongly non-parabolic in figure 1(b). Furthermore, from figures 2(e)-(h), we can find that the hole states are more localized than the electron states near the radial center $r$=0. When the diameter $D$ increases to 18 nm, the six lowest electron states and ten highest hole states are presented in figures 3(a) and 3(b), respectively. We can clearly see that the electron and hole states become more concentrated compared with the case of the diameter $D$=6 nm, because of the weakness of the quantum confinement effect. In figure 3(a), two groups of the electron states are very close, which have been enlarged in two small inset figures. The first excited state is not $e_1^{1/2}$, and becomes to $e_0^{3/2}$ at the $\Gamma$ point, whose main Bloch basis component is also $|S\rangle$$\uparrow$. However, in figure 3(b), it is found that the order of three highest hole states is unchanged.      

\begin{figure}[h!]
\centering\includegraphics[width=11cm, height=7cm]{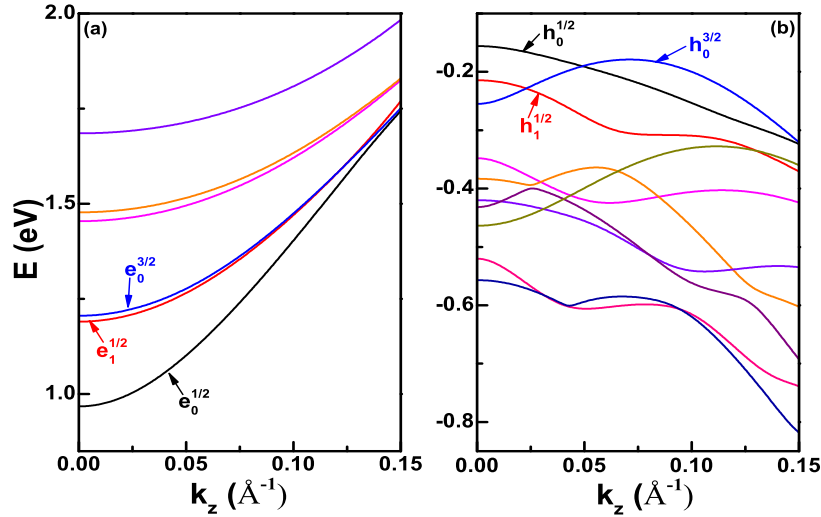}
\caption{(a) The six lowest electron states and (b) ten highest hole states of Ge nanowires with the diameter $D$=6 nm at the $\Gamma$-valley. }
\end{figure}
\begin{figure}[h!]
\centering\includegraphics[width=11cm, height=5.3cm]{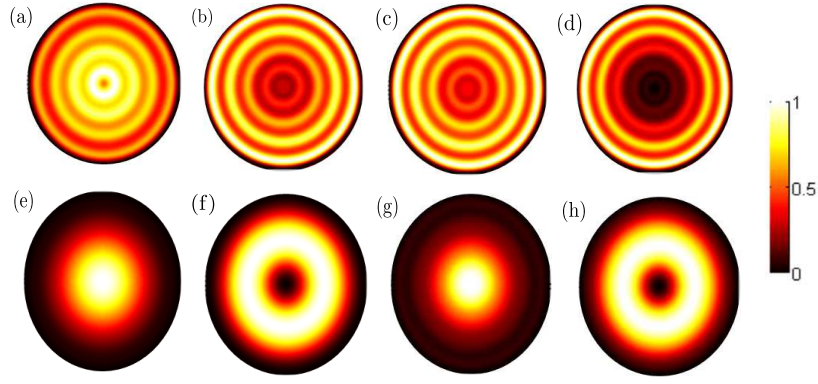}
\caption{The probability densities of the four lowest electron states and four highest hole states of Ge nanowires with the diameter $D$=6 nm at the $\Gamma$ point. (a)-(d) represent the electron states $e_0^{1/2}$, $e_1^{1/2}$, $e_0^{3/2}$ and $e_1^{3/2}$, respectively. (e)-(h) represent the hole states $h_0^{1/2}$, $h_1^{1/2}$, $h_0^{3/2}$ and $h_1^{3/2}$, respectively. }
\end{figure}
\begin{figure}[h!]
\centering\includegraphics[width=12cm, height=7cm]{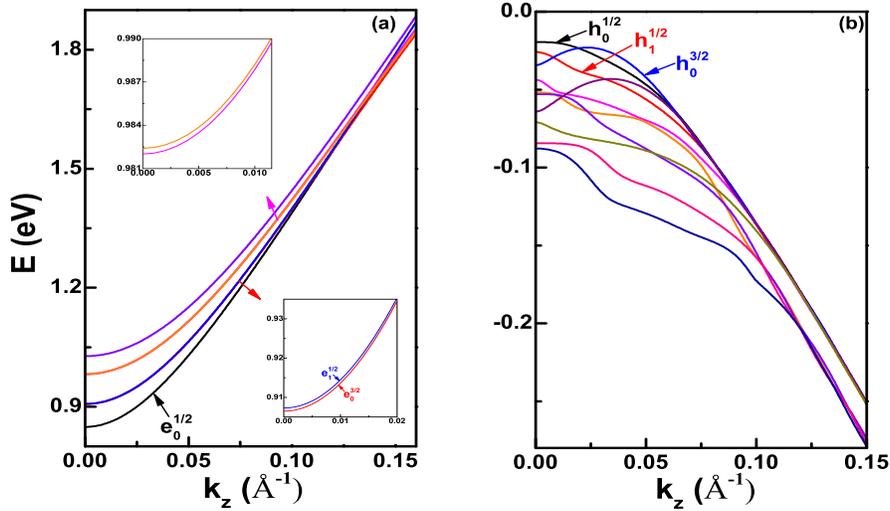}
\caption{(a) The six lowest electron states and (b) ten highest hole states of Ge nanowires with the diameter $D$=18 nm at the $\Gamma$-valley. The inset figures in (a) are the enlargement of two group close electron states, and the first group close electron state is $e_0^{3/2}$ and $e_1^{1/2}$.  }
\end{figure}

Figure 4(a) shows the six low electron states of Ge nanowires at the [111] $L$-valley when the diameter $D$ is 6 nm. As discussed above, each electron state is eight-fold degenerate in the Brillouin zone. However, at one paticular $L$-valley, each electron state is not degenerate for [001] Ge nanowires because the cylindrical symmetry is broken at the [111] $L$-valley, which is unlike the case at the $\Gamma$-valley. During the calculation, we must ensure the convergence of the energy $E_e^{L}$ by carefully choosing the appropriate cutting $m$ in Eq. (5). As seen in the figure, we find that not only the ground state but also the excited states are all almost parabolic. The term $k_{1}(k_{z}-\pi/a)$ in Eqs. (3) and (4) will couple $m$ and $m+1$, $m-1$ order Bessel functions, but the coupling is too weak in Ge nanowires to affect the parabolic behaviour of each electron state at the $L$-valley. Figures 4(b) and 4(c) are the same as figure 4(a) but for $D$=12 and 18 nm, respectively. Obviously, the electron states of these two figures are still approximately parabolic, while the curvature of the parabola of these electron states will increase as the increase of the diameter of Ge nanowires. In addition, we notice that the states become dense when $D$ increases from 6 nm to 18 nm, which is also due to the weakening of the quantum confinement effect. In figures 4(d)-4(g), the probabilities densities of four lowest electron states are plotted at the minimum $L$-valley, namely, $\pi/a$. Unlike the probability densities of the electron states at the minimum $\Gamma$-valley in figures 2(a)-(d), we can obviously see that the cylindrical symmetry of all these four electron states are broken at the minimum $L$-valley, and their probability densities depend on the azimuth angle $\theta$. Moreover, the ground electron state in figure 4(d) is most likely to appear near the center $r$=0, which means that $m$=0 term in Eq. (5) is the main contribution to its probability density. While two following electron states in figures 4(e) and 4(f) show very interesting probability density distribution, and $m$=1 and -1 terms in Eq. (5) primarily contributes to their probability densities.      
\begin{figure}[h!]
\centering\includegraphics[width=12cm, height=9cm]{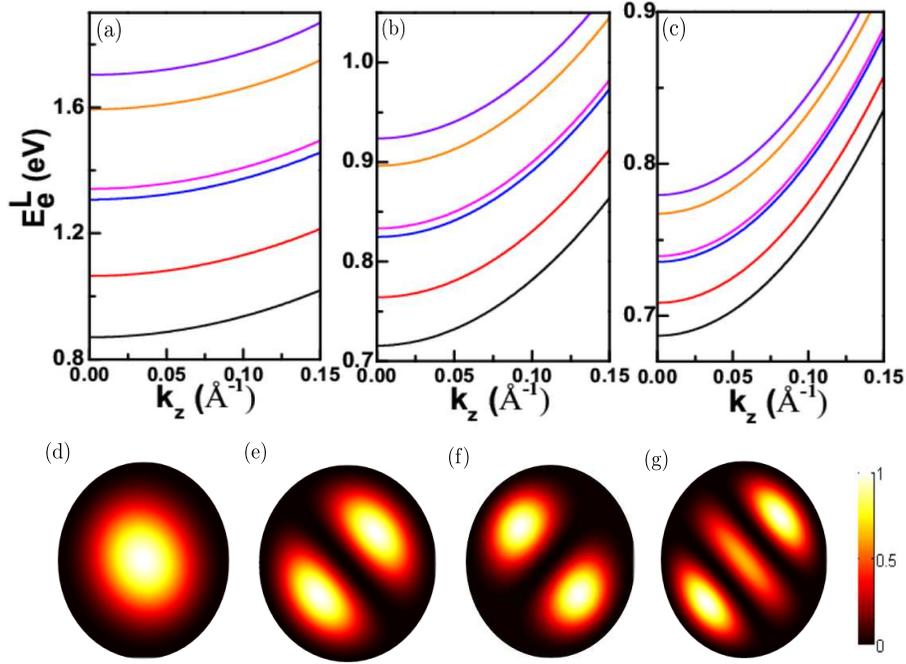}
\caption{(a) The six lowest electron states of Ge nanowirs with the diameter $D$=6 nm at the $L$-valley. (b) and (c) are the same as (a) but for the diameter $D$=12 nm and 18 nm, respectively. (d)-(g) are the probability densities of four lowest electron states of Ge nanowires with the diameter $D$=6 nm at the minimum $L$-valley, namely, $\pi/a$.  }
\end{figure}

\subsection*{The density of states and optical gain of Ge nanowires}

The density of states (DOS) are important for understanding the carrier occupation in Ge nanowires. Figure 5(a) shows the valence and conduction band DOS of Ge nanowires when the diameter $D$ is 6 nm, and the conduction band DOS with and without $L$-valley are presented in the figure for comparison. We can see that there are a series of sharp peaks in 
valence and conduction band DOS, which is a remarkable characteristic of one-dimensional density of states. Let us focus on the conduction band DOS. Obviously, the conduction band DOS will increase substantially when counting the electron states at the $L$-valley, which is due to a factor of 8 in the right side of Eq. (6). Therefore, the electrons will only 
\begin{figure}[h!]
\centering\includegraphics[width=14cm, height=7cm]{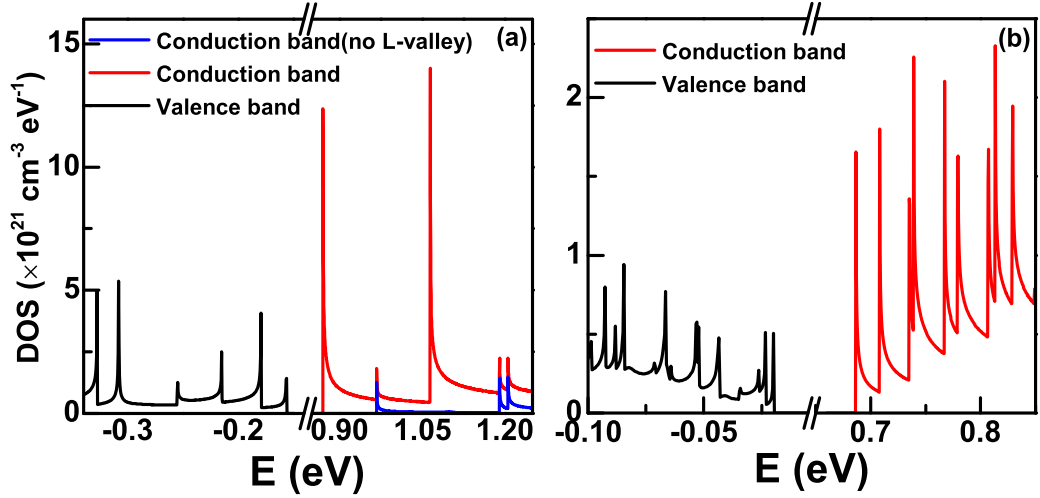}
\caption{(a) The conduction and valence band density of states (DOS) of Ge nanowires with the diameter $D$=6 nm. The conduction band DOS with and without $L$-valley are presented by red and blue solid line, respectively.  (b) is the same as (a) but for the diameter $D$=18 nm. The conduction band DOS including the $L$-valley is presented by red solid line.    }
\end{figure}
occupy the $L$-valley of Ge nanowires when the carrier concentration is low. As the increase of the concentration, a small amount of the electrons begin to fill the $\Gamma$-valley. Figure 5(b) shows the DOS of Ge nanowires when the diameter $D$=18 nm. We can see that more peaks will appear in the DOS as the increase of the diameter $D$, because the electron states become more and more dense, as shown in figures 1, 3 and 4.


As discussed in the above section, the quasi-Fermi levels $E_{fc}$ and $E_{fv}$ in Eqs. (8) and (9) can be determined for a given temperature and carrier concentration, thus we can further calculate the optical gain $g(E)$ of Ge nanowires via Eq. (10) according to their density of states. In the calculation, the refractive index and relaxation time are chosen as: $n_r$=4.02 and $\tau$=0.0658 ps\cite{32}, respectively, and the temperature is set to $T$=300 K. Meanwhile, the optical transition selection rules along $z$ or $x$ directions from all electron states to hole states should be satisfied. First of all, we consider the case that the doping concentration $N_d$ equals to 0. Figure 6(a) shows the optican gain spectra of Ge nanowires along $z$ direction as a function of the injected carrier density $\Delta n$ with $D$=6 nm when $N_d$=0. In the figure, $N_0$ represents the value 1$\times$$10^{19}$ cm$^{-3}$. We can see that there is no optical gain even though $\Delta n$=4 $N_0$, namely, 4$\times$$10^{19}$ cm$^{-3}$, which is due to nearly total occupation of the electrons in the $L$-valley. For example, when $\Delta n$=4 $N_0$, the occupied proportion of the $\Gamma$-valley is only about 0.385\%. Therefore, for indirect-band-gap Ge nanowires, we need to take some measures to improve the occupied proportion of the injected electrons at the $\Gamma$-valley in the future work. As the increase of $\Delta n$, more and more electrons begin to fill the $\Gamma$-valley, and the optical gain appears in the figure. Obviously, the peak value of the optical gain $g(E)$ can reach to about 7790.33 cm$^{-1}$ if  $\Delta n$ increases to 9 $N_0$. It can be analyzed that the peak of optical gain comes from the optical transition $e_0^{1/2}$$\Rightarrow$$h_0^{1/2}$ and $e_0^{-1/2}$$\Rightarrow$$h_0^{-1/2}$. Because $e_0^{1/2}$ and $e_0^{-1/2}$ are degenerate, so does $h_0^{1/2}$ and $h_0^{-1/2}$, these two optical transitions coincide at the photon 
\begin{figure}[h!]
\centering\includegraphics[width=11.5cm, height=12.5cm]{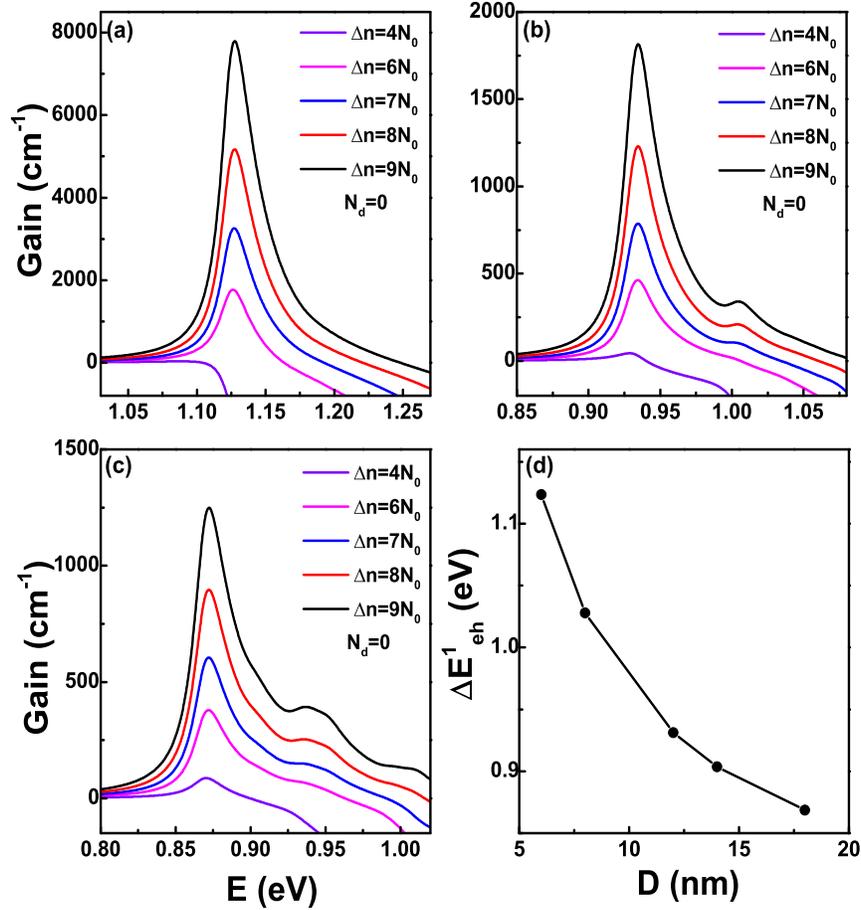}
\caption{(a) The optical gain spectra of Ge nanowires along $z$ direction as a fuction of the injected density $\Delta n$  with the diameter $D$=6 nm and the temperature $T$=300 K when the doping concentration $N_d$=0; (b) and (c) are the same as (a) but for the diameter $D$=12 nm and 18 nm, respectively; (d) The energy difference $\Delta$$E_{eh}^{1}$ at the $\Gamma$ point between the ground electron state and hole state as a function of the diameter $D$.   }
\end{figure}
energy $E$$\approx$1.127 eV, which belongs to the infrared range. The optical gain spectra of Ge nanowires as a function of $\Delta n$ is shown in figure 6(b) when the diameter
$D$=12 nm and $N_d$=0 . As seen in the figure, a obvious peak will appear as the increase of $\Delta n$, and the peak value of $g(E)$ can reach to about 1814.70 cm$^{-1}$ at the 
photon energy $E$$\approx$0.935 eV if $\Delta n$ is 9 $N_0$. In comparison with the case of the diameter $D$=6 nm in figure 6(a), the peak value of $g(E)$ decreases substantially. Further, the photon energy is still in the range of infrared range and shows the redshift compared with figure 6(a). Through the analysis, in figure 6(b), the first and large gain peak also originates from the optical transition $e_0^{1/2}$$\Rightarrow$$h_0^{1/2}$ and $e_0^{-1/2}$$\Rightarrow$$h_0^{-1/2}$, while a second and small gain peak will appear gradually as the increase of $\Delta n$. This small peak locates at the photon energy $E$$\approx$1.004 eV, and is from the optical transition $e_0^{1/2}$$\Rightarrow$$h_2^{1/2}$ and $e_0^{-1/2}$$\Rightarrow$$h_2^{-1/2}$. Figure 6(c) is the same as figures 6(a) and 6(b), respectively but for the diameter $D$=18 nm. In figure 6(c), the large peak value of $g(E)$ and the photon energy further decreases and shows redshift, respectively in comparison to the condition in figure 6(b), and similar to the cases in figures 6(a) and 6(b), we can also analyze the contribution of the optical transition to each peak. In figure 6(d), the energy difference $\Delta$$E_{eh}^1$ between the first electron state $e_0^{1/2}$ and the first hole state $h_0^{1/2}$ at the $\Gamma$ point as a function of the diameter $D$ is presented. We can see that  $\Delta$$E_{eh}^{1}$ decreases obviously as the increase of $D$ due 
to the reduction of quantum confinement effect, which also matches the redshift of energy corresponding to the peak gain in figures 6(a), 6(b) and 6(c). Therefore, according to the above analysis of the optical gain spectra along $z$ direction with three different diameters, we can conclude that there is almost no optical gain even though the injected carrier density $\Delta n$ is 4$\times$$10^{19}$ cm$^{-3}$ when $N_d$=0, which is due to the characteristics of the carrier preferential filling of Ge $L$-valley. As stated in previous section, the free-carrier absorption (FCA) loss must be taken into account for the heavily doped Ge nanowires. Figure 7(a) shows the peak gain along $z$ direction, free-carrier absorption (FCA) loss and net peak gain of Ge nanowires as a fuction of injected density $\Delta n$ with $D$=6 nm and $T$=300 K when $N_d$=0. From the figure, we can clearly see that FCA loss almost increases linearly as the increase of $\Delta n$, because more electrons will fill the $\Gamma$-valley, and the electron concentrations $N_{e}^{\Gamma}$ at the $\Gamma$-valley in formula (14) will increase if the injected density $\Delta n$ increases. FCA loss is greater than peak gain when $\Delta n$ is less than 5$\times$$10^{19}$ cm$^{-3}$, while as the further increase of $\Delta n$, FCA loss will become less than the peak gain, which means that the transparent injected density is in the range 4$\sim$5$\times$$10^{19}$ cm$^{-3}$ when $D$=6 nm and $N_d$=0. Figures 7(b) and 7(c) are the same as figure 7(a) but for the diameter $D$=12 nm and 18 nm, respectively. Obviously, although the peak gain will increase when $\Delta n$ increases from 4$\times$$10^{19}$ cm$^{-3}$ to 9$\times$$10^{19}$ cm$^{-3}$, the peak gain will always be less than FCA loss, which means that the negative net peak gain will be encountered when $N_d$=0 and $D$=12 nm or 18 nm. The reason is that the peak gain decreases sharply, while FCA loss change a little as the increase of $D$. We also analyze the optical gain along $x$ direction, and the results show that the first peak gain along $x$ direction with $D$=6 nm, 12 nm and 18 nm is all less than corresponding FCA loss even though $\Delta n$ increases to 9$\times$$10^{19}$ cm$^{-3}$.
\begin{figure}[h!]
\centering\includegraphics[width=15cm, height=7cm]{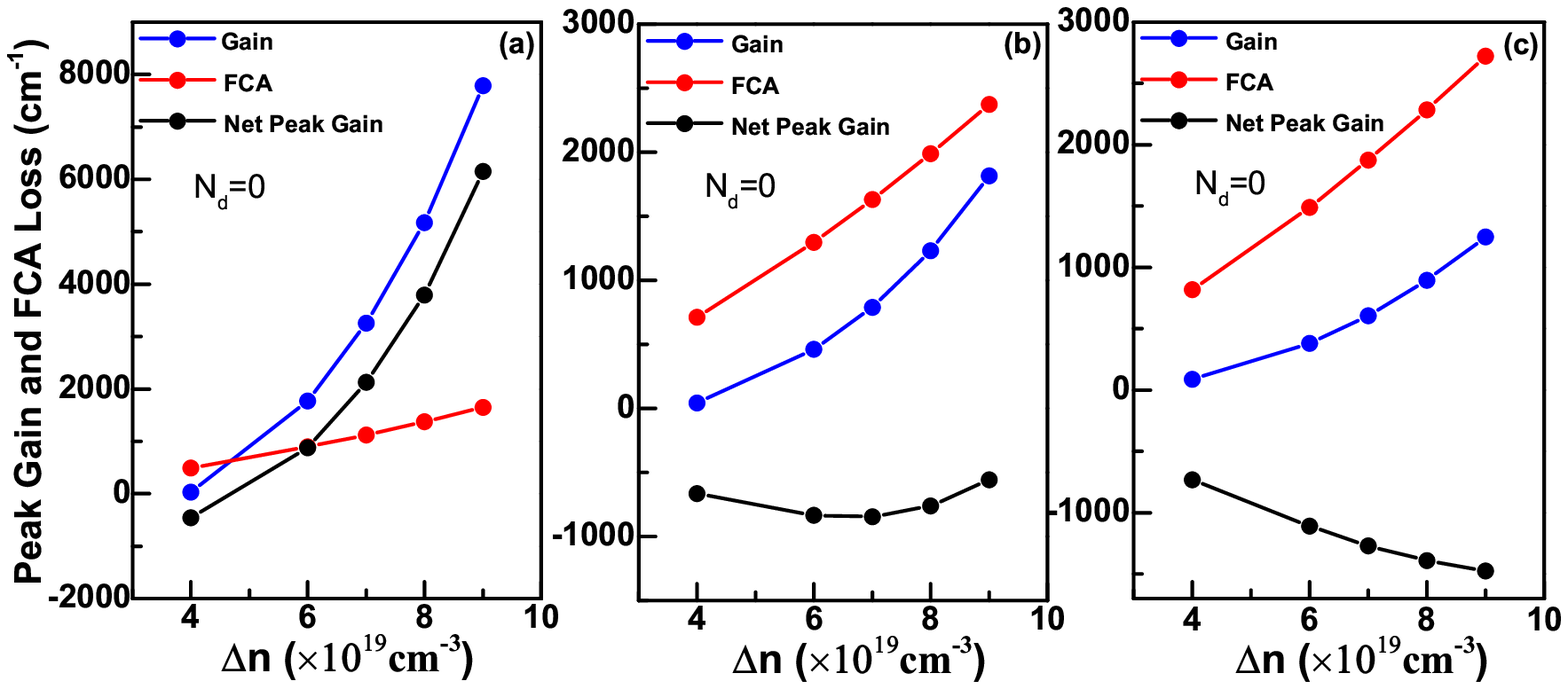}
\caption{(a) The peak gain along $z$ direction, free-carrier absorption (FCA) loss and net peak gain of Ge nanowires as a fuction of the injected density $\Delta n$ with the diameter $D$=6 nm and the temperature $T$=300 K when the doping concentration $N_d$=0; (b) and (c) are the same as (a) but for the diameter $D$=12 nm and 18 nm, respectively.      }
\end{figure}

\begin{figure}[h!]
\centering\includegraphics[width=15cm, height=7cm]{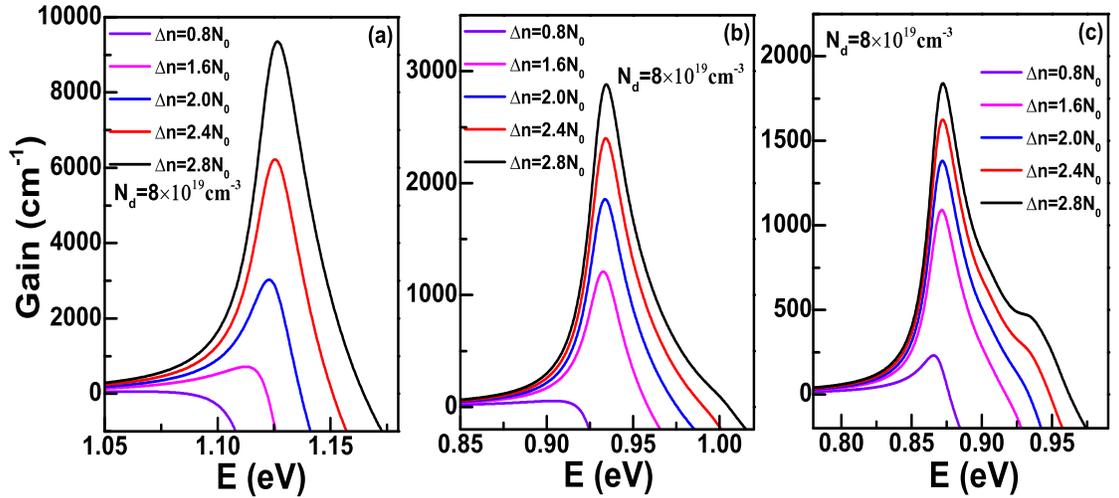}
\caption{(a) The optical gain spectra of Ge nanowires along $z$ direction as a fuction of the injected density $\Delta n$  with the diameter $D$=6 nm and the temperature $T$=300 K when the doping concentration $N_d$=8$\times$$10^{19}$ cm$^{-3}$; (b) and (c) are the same as (a) but for the diameter $D$=12 nm and 18 nm, respectively.     }
\end{figure}

\begin{figure}[h!]
\centering\includegraphics[width=15cm, height=7cm]{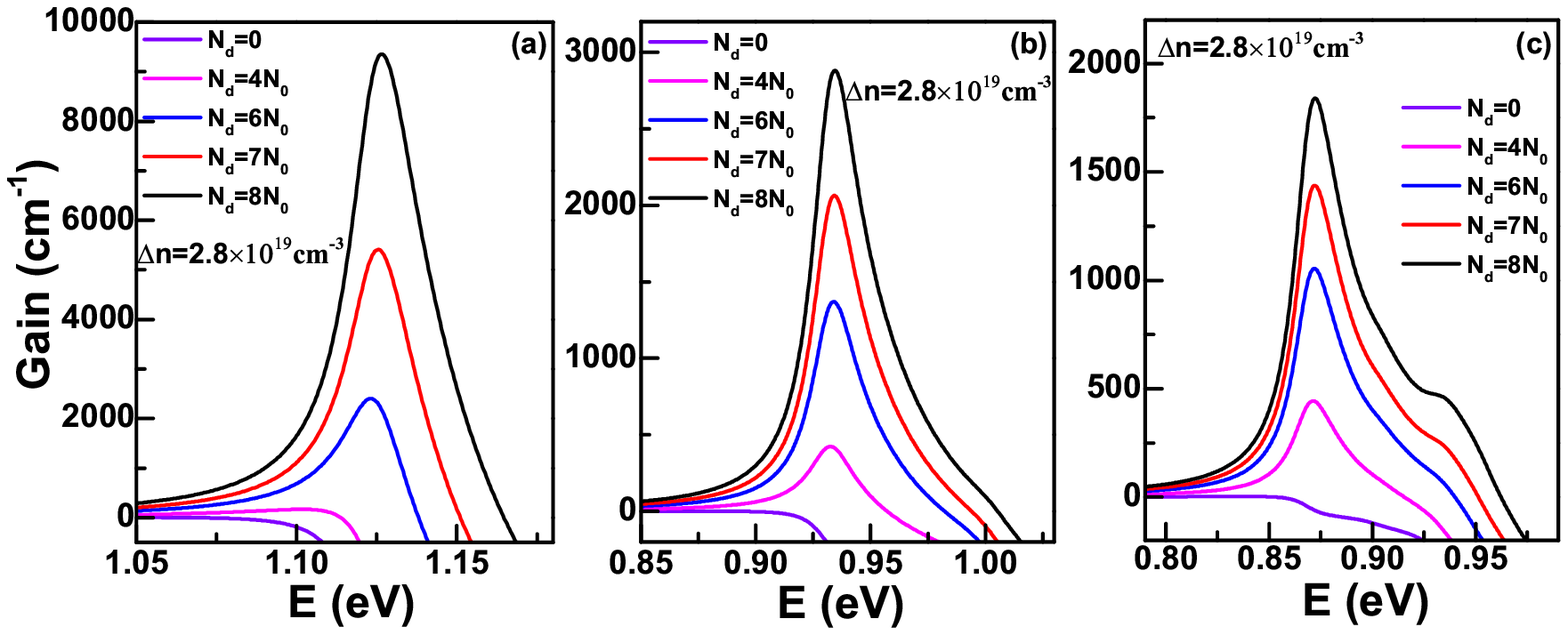}
\caption{(a) The optical gain spectra of Ge nanowires along $z$ direction as a fuction of the doping concentration $N_d$ with the diameter $D$=6 nm and the temperature $T$=300 K when the injected density $\Delta n$=2.8$\times$$10^{19}$ cm$^{-3}$; (b) and (c) are the same as (a) but for the diameter $D$=12 nm and 18 nm, respectively.     }
\end{figure}
 
Secondly, we will discuss the case that the doping concentration $N_d$ is not equal to 0. Figure 8(a) demonstrates the optical gain along $z$ direction as a function of the injected density $\Delta n$ with $D$=6 nm when $N_d$ is fixed to 8$\times$$10^{19}$ cm$^{-3}$. It can be seen that the optical gain is nearly zero when $\Delta n$=0.8$\times$$10^{19}$ cm$^{-3}$ and will become larger than 0 from  $\Delta n$=1.6$\times$$10^{19}$ cm$^{-3}$. When $\Delta n$ increases to 2.8$\times$$10^{19}$ cm$^{-3}$, the peak gain is about 9355.15 cm$^{-1}$ at the photon energy $E$$\approx$1.1264 eV, which corresponds to the optical transition $e_0^{1/2}$$\Rightarrow$$h_0^{1/2}$. The optical gain along $z$ direction with $D$=12 and 18 nm as a function of $\Delta n$ are also presented in figures 8(b) and 8(c). In figure 8(b), we find that the peak gain is about 2880.75 cm$^{-1}$ at the photon energy $E$$\approx$0.9343 eV when $\Delta n$ increases to 2.8$\times$$10^{19}$ cm$^{-3}$, while the peak gain is about 1840.00 cm$^{-1}$ with the largest $\Delta n$ in figure 8(c), which is less than the corresponding peak gain with the same injected carrier density in figure 8(b). Furthermore, we also calculate the optical gain along $z$ direction as a function of $N_d$ when $\Delta n$ is fixed, and the results are displayed in figure 9. In figure 9(a), the peak gain is very small when $N_d$ is 4$\times$$10^{19}$ cm$^{-3}$, then the peak gain increases rapidly as the increase of  $N_d$, which can reach to about 9355.15 cm$^{-1}$ if $N_d$ is 8$\times$$10^{19}$ cm$^{-3}$. In figures 9(b) and 9(c), the diameter $D$ are 12 nm and 18 nm, respectively, we can find that the peak gain will increase slowly as the increase of $N_d$. Considering FCA loss, the net peak gain as functions of the diameter $D$ and doping concentration $N_d$ is domenstrated in figure 10 when the injected carrier density $\Delta n$=2.8$\times$$10^{19}$ cm$^{-3}$ is fixed. We can see that there is a positive net peak gain with $D$=6 nm when $N_d$ is slightly larger than 5$\times$$10^{19}$ cm$^{-3}$, and the net peak gain can increase to about 8100.20 cm$^{-1}$ as the increase of  $N_d$. However, as the increase of $D$, the transparent doping concentration will increase. It can be clearly found that the transparent doping concentration can be up to about 6.5$\times$$10^{19}$ cm$^{-3}$ as $D$ increases to 14 nm, while if $D$=18 nm, the net peak gain becomes negative even though $N_d$ is 8$\times$$10^{19}$ cm$^{-3}$, because in this case, the peak gain decreases to about 1840.00 cm$^{-1}$, which is smaller than the corresponding FCA loss. Therefore, from figure 10, we can understand that it is getting harder to obtain positive net optical gain along $z$ direction as the increase of $D$, and the reason is that the peak gain will decrease as the increase of $D$ when the doping concentration $N_d$ and injected carrier density $\Delta n$ keep the same. Let us recall figures 7(b) and 7(c), when the doping concentration $N_d$=0, the negative net peak gain is encountered no matter what the carrier density is injected when the diameter $D$ is 12 nm or 18 nm. While in figure 10, we can obtain positive peak gain even though the diameter $D$ is larger than 12 nm. Because in the latter case, the hole concentration is equal to $\Delta n$, which is smaller than that of the former case. In addition,  FCA loss relates to the hole concentration, as shown in formula (14), which causes the FCA loss of the latter case to be smaller than that of the former case, thus a positive net peak gain is more likely to occur in the latter case.     

\begin{figure}[h!]
\centering\includegraphics[width=11cm, height=9cm]{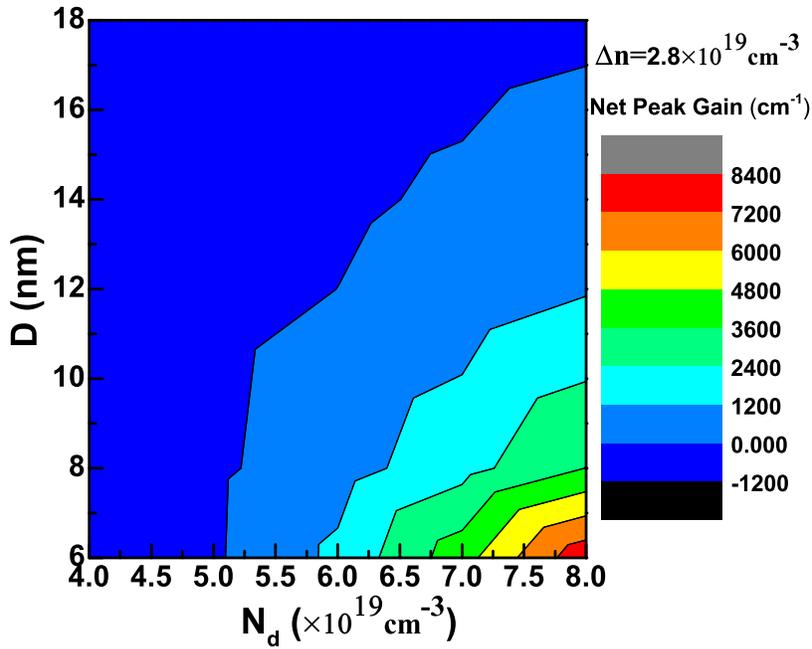}
\caption{(a) The net peak gain of Ge nanowires along $z$ direction as fuctions of the doping concentration $N_d$ and the diameter $D$ when the injected density $\Delta n$=2.8$\times$$10^{19}$ cm$^{-3}$ is fixed.    }
\end{figure}

\section*{Summary}

In a word, the electronic structures of the indirect-band-gap Ge nanowires are calculated via the effective-mass theory. Especially, for the first time, the energies of the electron states at the [111] $L$-valley are solved by using the Bessel function expansion method, which is important for understanding the optical properties of Ge nanowires because the injected electrons will fill the $L$-valley firstly. The calculated results show that the energy-wave vector dispersion relation at the $L$-valley is almost parabolic regardless of the diameter is 6 nm or 18 nm. Further, the density of states and direct-band-gap optical gain along $z$ direction are calculated on account of the electron and hole states at the $\Gamma$-valley and $L$-valley. We find that, due to the eight-fold degeneracy of each state at the $L$-valley, the conduction band DOS is mainly from the $L$-valley. The calculations of the optical gain spectra show that the gain peak locates in the infrared range, and there is almost no optical gain even though the injected carrier density is 4$\times$$10^{19}$ cm$^{-3}$ when the doping concentration is zero. In this case, the negative optical gain will be encountered considering FCA loss of Ge nanowires as the increase of the diameter. Meanwhile, the optical gain along $z$ direction as functions of the doping concentration and injected carrier density are also calculated when the doping concentration is not zero, and FCA loss is also taken into account. The results show that a positive net peak gain is most likely to occur in the heavily doped Ge nanowires with smaller diameters. Our theoretical studies are helpful for the applications of Ge nanowires in the field of microelectronics and optoelectronics.   

\section*{Appendix}

In the Hamiltonian $H_L^{[111]}$ at the [111] $L$-valley, we should express the operators $k_1$ and $k_2$ in following cylindrical coordinate because the nanowires is confined in $(r,\theta)$ plane. 
\begin{equation}
\begin{aligned}
 k_1=\frac{1}{\sqrt{2}}(k_x+k_y) & =-\frac{i}{\sqrt{2}}(\frac{\partial}{\partial x}+\frac{\partial}{\partial y}) \\ &  = -\frac{i}{\sqrt{2}}\Big[\big(cos(\theta)+sin(\theta)\big)\frac{\partial}{\partial r}+\frac{\big(cos(\theta)-sin(\theta)\big)}{r}\frac{\partial}{\partial \theta}\Big] 
\end{aligned}
\end{equation}
\begin{equation}
\begin{aligned}
 k_2=\frac{1}{\sqrt{2}}(-k_x+k_y) & =-\frac{i}{\sqrt{2}}(-\frac{\partial}{\partial x}+\frac{\partial}{\partial y}) \\ &  = -\frac{i}{\sqrt{2}}\Big[\big(-cos(\theta)+sin(\theta)\big)\frac{\partial}{\partial r}+\frac{\big(cos(\theta)+sin(\theta)\big)}{r}\frac{\partial}{\partial \theta}\Big]
\end{aligned}
\end{equation}
Therefore, the operators $k_1^2$ and $k_2^2$ can be written as
\begin{equation}
\begin{aligned}
k_1^2= & -\frac{1}{2}\Big[\big(1+sin(2\theta)\big)\frac{\partial^2}{\partial^2 r}+2cos(2\theta)\big(\frac{1}{r}\frac{\partial^2}{\partial r\partial \theta}-\frac{1}{r^2}\frac{\partial}{\partial \theta}\big)+ \\ &
\big(1-2sin(2\theta)\big)\big(\frac{1}{r}\frac{\partial}{\partial r}+\frac{1}{r^2}\frac{\partial^2}{\partial^2 \theta}\big)\Big]
\end{aligned}
\end{equation}
\begin{equation}
\begin{aligned}
k_2^2= & -\frac{1}{2}\Big[\big(1-sin(2\theta)\big)\frac{\partial^2}{\partial^2 r}-2cos(2\theta)\big(\frac{1}{r}\frac{\partial^2}{\partial r\partial \theta}-\frac{1}{r^2}\frac{\partial}{\partial \theta}\big)+ \\ &
\big(1+2sin(2\theta)\big)\big(\frac{1}{r}\frac{\partial}{\partial r}+\frac{1}{r^2}\frac{\partial^2}{\partial^2 \theta}\big)\Big]
\end{aligned}
\end{equation}
We can see that there are trigonometric functions $sin(2\theta)$ and $cos(2\theta)$ in the operators $k_1^2$ and $k_2^2$, which will couple $m-2$, $m$ and $m+2$ order Bessel functions because the following relations
\begin{equation}
sin(2\theta)=\frac{1}{2i}\big(e^{2i\theta}-e^{-2i\theta}\big), ~~~~ cos(2\theta)=\frac{1}{2}\big(e^{2i\theta}+e^{-2i\theta}\big)    \\
\end{equation}
Similarly, there is trigonometric functions $sin(\theta)$ and $cos(\theta)$ in the operators $k_1$ and $k_2$, which will couple $m-1$, $m$ and $m+1$ order Bessel functions. Obviously, there are three parts in the operator $k_1^2$, namely, 
\begin{equation}
P_1=-\frac{1}{2}\big(1+sin(2\theta)\big)\frac{\partial^2}{\partial^2 r}
\end{equation}
\begin{equation}
P_2=-cos(2\theta)\big(\frac{1}{r}\frac{\partial^2}{\partial r\partial \theta}-\frac{1}{r^2}\frac{\partial}{\partial \theta}\big)
\end{equation}
and 
\begin{equation}
P_3=-\frac{1}{2}\big(1-2sin(2\theta)\big)\big(\frac{1}{r}\frac{\partial}{\partial r}+\frac{1}{r^2}\frac{\partial^2}{\partial^2 \theta}\big)
\end{equation}
we must calculate the matrix elements $\langle n',m'|P_1|n,m\rangle$, $\langle n',m'|P_2|n,m\rangle$ and $\langle n',m'|P_3|n,m\rangle$, where $|n,m\rangle$=$A_{n,m}j_m(k_n^{m}r)e^{im\theta}$, $m$ and $m'$ denote the order of the Bessel functions, and $n$ and $n'$ denote the zero points of the Bessel functions. 

First of all, the matrix elements $\langle n',m'|P_1|n,m\rangle$ can be expressed as 
\begin{equation}
\langle n',m'|P_1|n,m\rangle=\frac{i}{16}2\pi R^2(k_n^m)^2(I_1)\big(\delta_{m,m'-2}-\delta_{m,m'+2}+2i\delta_{m,m'}\big)
\end{equation}
where 
\begin{equation}
I_1=A_{n',m'}A_{n,m}\int_0^{1}j_{m'}(\alpha_{n'}^{m'}r)\big[j_{m-2}(\alpha_{n}^{m}r)+j_{m+2}(\alpha_{n}^{m}r)-2j_{m}(\alpha_{n}^{m}r)\big]rdr
\end{equation}
Secondly, the matrix elements $\langle n',m'|P_2|n,m\rangle$ can be expressed as
\begin{equation}
\langle n',m'|P_2|n,m\rangle=\big[-\frac{i}{4}2\pi Rm (k_n^{m})(I_2)+\frac{i}{2}2\pi m(I_3)\big]\big(\delta_{m,m'-2}+\delta_{m,m'+2}\big)
\end{equation}
where 
\begin{equation}
I_2=A_{n',m'}A_{n,m}\int_0^{1}j_{m'}(\alpha_{n'}^{m'}r)\big[j_{m-1}(\alpha_{n}^{m}r)-j_{m+1}(\alpha_{n}^{m}r)\big]dr
\end{equation}
\begin{equation}
I_3=A_{n',m'}A_{n,m}\int_0^{1}j_{m'}(\alpha_{n'}^{m'}r)j_{m}(\alpha_{n}^{m}r)\frac{dr}{r}
\end{equation}
Finally, the matrix elements $\langle n',m'|P_3|n,m\rangle$ can be expressed as
\begin{equation}
\langle n',m'|P_3|n,m\rangle=\big[\frac{i}{8}2\pi R (k_{n}^{m})(I_2)-\frac{i}{4}2\pi m^{2}(I_3)\big]\big(-\delta_{m,m'-2}+\delta_{m,m'+2}+2i\delta_{m,m'}\big) 
\end{equation}
During the calculation of the above three matrix elements, the following relation of the Bessel function is used
\begin{equation}
\frac{dj_m(k_n^{m}r)}{dr}=\frac{1}{2}\big(k_n^m\big)\big[j_{m-1}(k_n^{m}r)-j_{m+1}(k_n^{m}r)\big]
\end{equation}

Apart from the operator $k_1^2$, there are two operators $k_2^2$ and $k_1$ in Hamiltonian $H_L^{[111]}$. Obviously, the expression of the operator $k_2^2$ is very close to that of the operator $k_1^2$, and we can infer the matrix elements of $k_2^2$. The matrix elements of the operator $k_1$ can be expressed as
\begin{equation}
\begin{aligned}
\langle n',m'|k_1|n,m\rangle= & -2\pi R^{2}(k_n^{m})(I_4)\big[\frac{(1+i)}{4\sqrt{2}}\delta_{m,m'-1}+\frac{(-1+i)}{4\sqrt{2}}\delta_{m,m'+1}\big]+ \\ & 2\pi Rm(I_5)\big[\frac{(1+i)}{2\sqrt{2}}\delta_{m,m'-1}+\frac{(1-i)}{2\sqrt{2}}\delta_{m,m'+1}\big] 
\end{aligned}
\end{equation}
where
\begin{equation}
I_4=A_{n',m'}A_{n,m}\int_0^{1}j_{m'}(\alpha_{n'}^{m'}r)\big[j_{m-1}(\alpha_{n}^{m}r)-j_{m+1}(\alpha_{n}^{m}r)\big]rdr
\end{equation}
\begin{equation}
I_5=A_{n',m'}A_{n,m}\int_0^{1}j_{m'}(\alpha_{n'}^{m'}r)j_{m}(\alpha_{n}^{m}r)dr
\end{equation}
Finally, the energies of the electron states at the [111] $L$-valley can be solved numerically. 


\section*{Acknowledgements (not compulsory)}

This work is funded by National Research Foundation of Singapore (NRF-CRP19-2017-01) and Fundamental Research
Funds for the Central Universities (2018CDXYWU0025).

\section*{Author contributions statement}

W. J. Fan and C. S. Tan suggested the project direction; W. Xiong and J. W. Wang performed the theoretical calculations and analyzed the results; Z. G. Song took part in the discussions and put forward constructive opinions. W. Xong organized and wrote the paper. All authors reviewed and approved the manuscript. 
 
\section*{Additional information}

Competing Interests: The authors declare no competing interests. \\
Publisher’s note: Springer Nature remains neutral with regard to jurisdictional claims in published maps and
institutional affiliations

\end{document}